\newif\iffiginc
\figincfalse  
\iffiginc
\documentstyle[prd,aps,preprint,tighten,floats,psfig]{revtex}
\else
\documentstyle[prd,aps,preprint,tighten,floats]{revtex}
\fi
\begin{document}

\preprint{\bf UG--FT--33/93\hskip0.5cm
UPR--590--T}
\date{\bf December 1993}

\title{
   Diagnostic Power of  Future Colliders  for ${\bf Z^\prime}$
   Couplings to Quarks and
Leptons: ${\bf e^+e^-}$ versus ${\bf pp}$  Colliders}

\author{F. DEL AGUILA}
\address{Departamento de F\'\i sica Te\'orica y del Cosmos,
Universidad de Granada \\
Granada, 18071, Spain\\
\ \ \ \\
{\rm and}}

\author{M. CVETI\v C}
\address{Department of Physics, University of Pennsylvania \\
Philadelphia, PA 19104-6396}

\maketitle

\begin{abstract}
We study the diagnostic power of  future  $e^+e^-$ colliders with ${\sqrt {s}}
= 500$
GeV (the New Large Collider ($NLC$)) for a model independent determination of
the
  $Z'$ gauge couplings to quarks and leptons.  The interference of the  $Z'$
propagator
with the photon and the $Z$ propagator   in the  two-fermion  final state
probes are sensitive to  the  magnitude  as well as   relative signs of
quark and lepton charges.  For
$Z'$  with $M_{Z'} \sim 1 $  TeV  {\it all } the  quark and  lepton charges
can be determined to around $10-20 \%$, provided heavy flavor tagging
and   longitudinal polarization of the  electron beam is available.  The
errors are $2-10$ times larger without polarization, and very little
information can be
obtained  about quark charges without heavy flavor tagging.
We point out   the complementarity of future hadron colliders.
At the  CERN  Large Hadron Collider ($LHC$) primarily the magnitude of
three out of four  corresponding   couplings  can be measured; however,
their error-bars are  typically by a factor of $\sim 2$ smaller than those
at  the $NLC$.
\end{abstract}

\section{Introduction}

If  the masses of  heavy gauge bosons $Z'$'s
 do not exceed $5$ TeV or so,   the CERN Large Hadron Collider ($LHC$),
would be  an ideal place to discover them \cite{ACLI}\ .
In the last few years a number of diagnostic probes have been proposed
\cite{ACLI}\ , allowing  for a model independent  determination\cite{ACLII}\
of certain $Z^\prime$ couplings
to quarks and leptons provided
$M_{Z'}{\lower3pt\hbox{$\buildrel<\over\sim$}} 2$ TeV.

On the other hand, future $e^+e^-$ colliders
 with large enough center of mass energy $\sqrt s$ , {\it e.g.}, $\sqrt{ s} =2$
TeV, could provide a clean way to discover and study the properties of $Z'$'s.
A more likely possibility, however, is the next linear collider ($NLC$)  with
$\sqrt{s}=500$ GeV.
Due to the interference of the $Z^\prime$ propagator with the photon  and
$Z$  propagators, the  two-fermion  channels
yield complementary   information on
the existence of a  $Z'$.  An extensive study \cite{DLRSV,HR}\ showed that
effects of a $Z'$ would be observable at the $NLC$ for  a large class of models
with $M_{Z'}$ up to  $1-3$ TeV.  In particular, in Ref. \cite{DLRSV}\
the sensitivity of the $NLC$  to specific classes of
 extended electroweak models, {\it e.g.},
different $E_6$ motivated models described by a parameter $\cos \beta$
(the mixing  angle between the $Z_\chi$  and $Z_\psi$ defined below) or
left-right symmetric models parameterized by the ratio $\kappa=g_R/g_L$ for
the $SU(2)_{L,R}$ gauge coupling constants $g_{L,R}$,
was explored.

In this paper we   explore  further the diagnostic power of the
$NLC$  for  $Z^\prime$  physics. In particular,  we investigate
a model independent  determination of the $Z^\prime$ couplings to quarks and
leptons \cite{FOOTB} .  We take the attitude that at
the $LHC$, which  is likely to be built before the $NLC$, $Z'$ would
either be discovered  or  strong bounds on  $M_{Z'}$ ($>$  5 TeV for
typical classes of models)
would  be achieved. Only in the former case   would the $NLC$
provide a testing ground to learn more about the  $Z'$.
We  therefore assume that the $Z'$  has a mass in the  range of a few
TeV, and  thus  the $NLC$  has the  capability to probe the $Z'$ couplings.

 We shall see
that  heavy flavor ($c,b,t$) tagging  would provide  a crucial diagnostic tool
for the
determination of the quark couplings.
 Based on   the  success of LEP experiments in  measuring quark cross sections
for different heavy flavors ($c,b$)\cite{LEP}\ , we will   assume that
 heavy flavor tagging  will be feasible at the $NLC$.
Another crucial tool is the longitudinal polarization of the
electron beam, which turns out to be important  for an unambiguous
determination of the  lepton couplings, including their relative signs.
Heavy flavor tagging, along with the  longitudinal
polarization of the electron beam,
 provide probes in the two-fermion final state channels which are sensitive
 to  the  magnitude as well as the relative signs of  {\it all} the $Z'$
charges
to quarks  and leptons. It turns out that for $M_{Z'} \sim$ 1 TeV, such
couplings would  be determined
to about $10-20$\% at the $NLC$.
If polarization were not    available,
the determination of the  $Z'$  couplings  would  be  marginal, since the
error-bars
increase by a factor of $2-10$. Similarly,  without heavy flavor tagging, very
little  can be learned about the quark couplings.

Another goal of this paper is to   compare the analysis done for  the $NLC$
with
 the one that has been done for  the  $LHC$ collider\cite{ACLII}\ .
The diagnostic power of the  $LHC$ is complementary.
It  allows primarily for the  determination  of  the magnitude
of three out of four normalized couplings, only.
However, the corresponding error-bars   are typically by a factor of $\sim
  2$
 smaller than  those for the
$NLC$.  In addition, the $LHC$ would measure $M_{Z'}$  directly and would
allow for a determination of an overall strength of the $Z'$ gauge coupling to
fermions. This is  in contrast to  the $NLC$ which, for fixed c.m.
energy, primarily determines only the  ratio of an overall $Z'$ gauge coupling
strength
and $M_{Z'}$.

The paper is organized as follows. In Section II we specify the notation
and the models used to illustrate the analysis. In Section III we discuss the
probes
for the two--fermion final state channels at the $NLC$. In Section IV simulated
 fits for  the $Z'$ charges to quark and leptons  are performed  for a class of
typical models.
In Section V we  compare results at the $NLC$  with
 those at the  $LHC$.  Conclusions are given in Section VI.

\section{Typical Models and  $Z'$ Couplings}

The neutral current gauge interaction term in the presence of an
additional $U_1$ is of the form \cite{LL}
\begin{equation}
-L_{NC}=eJ_{em}^\mu A_\mu +g_1 J_1^\mu Z_{1\mu}+ g_2J_2^\mu Z_{2\mu},
\end{equation}
with $Z_1$ the $SU_2 \times U_1$ boson and $Z_2$ the additional
boson in the weak eigenstate basis.  Here
$g_1\equiv\sqrt{g_L^2+g_Y^2}=g/\cos\theta_W$, where
 $g_L$, $g_Y$ are the
gauge couplings of $SU_{2L}$ and $U_{1Y}$,
and $g_2$ is the gauge coupling of $Z_2$.
The currents   are:
$J_j^\mu={1\over2}\sum_i{\bar\psi_i}\gamma^\mu
\left[\hat{g}^i_{V_j}-\hat{g}^i_{A_j}\gamma_5\right] \psi_i,
\quad j=1,2,
$
where the sum runs over fermions, and the $\hat{g}^i_{(V,A)_j}$
are the vector and axial vector couplings of $Z_j$ to the
$i^{th}$ flavor.  Analogously, $\hat{g}^i_{(L,R)_j}={1\over2}(
\hat{g}^i_{V_j}\pm \hat{g}^i_{A_j})$.

For illustration we consider the following typical
 GUT, left-right  symmetric, and
superstring-motivated models.

\begin{itemize}
\item $\chi$ model: $Z_\chi$ occurs in $SO_{10}\rightarrow SU_5\times
U_{1\chi}$.
\item$\psi$ model: $Z_\psi$ occurs in $E_6\rightarrow SO_{10}\times U_{1\psi}$.
 \item$\eta$ model: $Z_\eta=\sqrt{3/8}Z_\chi-\sqrt{5/8}Z_\psi$ occurs in
superstring inspired models in which $E_6$ breaks directly to a rank 5
group.
\item$LR$ model: $Z_{LR}$ occurs in left-right  (LR) symmetric models. Here
we consider the special value
$\kappa=g_R/g_L=1 $ of the gauge couplings $g_{L,R}$ for $SU_{2L,2R}$,
respectively.
\end{itemize}

In the rest of the paper we assume family universality
 and neglect $Z-Z'$
mixing (as  suggested from  experiments).
We also assume $[Q',T_i]=0$,
where $Q'$ is the $Z'$ charge  and $T_i$ are the $SU_{2L}$ generators,
which holds for a large class of models, including  the above
 $SU_2 \times U_1 \times U_1'$ and LR models.
The relevant quantities  to
 distinguish  between different models are  then the five
 charges: $\hat{g}^u_{L2}=\hat{g}^d_{L2}\equiv\hat{g}^q_{L2}$,
$\hat{g}^u_{R2}$, $\hat{g}^d_{R2}$, $\hat{g}^\nu_{L2}=\hat{g}^e_{L2}
\equiv\hat{g}^\ell_{L2}$, and $\hat{g}^\ell_{R2}$, and the gauge
coupling strength $g_2$.  The overall scale of the charges (and $g_2$)
depends on the normalization convention for ${\hbox{Tr}}(Q'^2)$,
but the ratios characterize particular theories.

Note that one combination of the five charges can always be absorbed in the
redefinition of   an overall
gauge coupling strength.   Since the photon couplings are only vector--like
and
the   $\ell$ couplings to $Z$ have the property
$\hat g_{L1}^\ell\simeq -\hat g_{R1}^\ell$  it turns out that
the  probes in the two--fermion
final state channels single out the $Z'$ leptonic couplings primarily in the
combinations $
\hat g_{L2}^\ell\pm\hat g_{R2}^\ell$.  To trace the
combinations of the normalized charges to which the  probes are sensitive, it
is
advantageous to choose either of the two  combinations to normalize
the charges.
 We choose the ${\hat g^\ell_{L2} - \hat g^\ell_{R2}}$ combination,
 which turns out to be a convenient choice for the typical models used in the
analysis. We then define  the following { four}
independent ``normalized''  charges:

\begin{equation}
P_V^\ell = \frac {\hat g^\ell_{L2} + \hat g^\ell_{R2}}
{\hat g^\ell_{L2} - \hat g^\ell_{R2}},
\ P_L^q = \frac {\hat g^q_{L2}}
{\hat g^\ell_{L2} - \hat g^\ell_{R2}},
\ P_R^{u,d} = \frac {\hat g^{u,d}_{R2}}
{\hat g^q_{L2}}.
\end{equation}

\noindent
Their values  are given for the typical models in Table I.
In addition, the probes in the   two-fermion final state channels
 are sensitive to  the  following ratio of  an  overall gauge coupling strength
divided by the ``reduced'' $Z'$ propagator:
\begin{equation}
\epsilon_A= (\hat g_{L2}^\ell
- \hat g_{R2}^\ell)^2 \frac {g_2^2}{4\pi \alpha }
\frac {s}{M^2_{Z'} - s}.
\end{equation}
 Here $\alpha$ is the fine structure constant.
Note again that the four normalized charges  (Eq.(2))
and $\epsilon_A$ (Eq.(3))
can be replaced with an equivalent set  by choosing
  ${\hat g^\ell_{L2} +\hat g^\ell_{R2}}$ to normalize the
couplings.

One  should contrast the  above choice of the normalized couplings  with those
chosen for the $LHC$. There the signs of the couplings are difficult to
determine and the following set of  four normalized
couplings is probed directly\cite{ACLII}\ :

\begin{equation}
\ \gamma _L^\ell\equiv \frac {{(\hat g^\ell_{L2})}^2 }
{{(\hat g^\ell_{L2})}^2 + {(\hat g^\ell_{R2})}^2}\ ,\ \
 \gamma _L^q \equiv\frac {{(\hat g^q_{L2})}^2 }
{{(\hat g^\ell_{L2})}^2 + {(\hat g^\ell_{R2})}^2}\ ,\ \
\tilde U\equiv \frac {{(\hat g_{R2}^u)}^2}{{(\hat g^q_{L2})}^2 }\ ,
\ \tilde D \equiv
\frac {{(\hat g_{R2}^d)}^2}{{(\hat g^q_{L2})}^2 }\ ,
\end{equation}
 which  can be expressed in terms of the couplings (2) as:
\begin{equation}
\ \gamma _L^\ell=\frac {(1 + P_V^\ell)^2}{2(1 + P_V^{\ell\ 2})},
\\ \  \gamma _L^q = \frac {2P_L^{q\ 2}}{1 + P_V^{\ell\ 2}},\ \
\break \tilde U= {(P_R^{u})}^2 ,\  \ \tilde D ={( P_R^{d})}^2.
\end{equation}
In addition, for $M_{Z'}
{\lower3pt\hbox{$\buildrel<\over\sim$}} 5$ TeV, the $LHC$  would  determine
$M_{Z'}$ and the total width $\Gamma_{Z'}$ directly in the main discovery
channel $pp\rightarrow Z'\rightarrow \ell^+\ell^-$ ($\ell=e,\mu$).
 Then
the  quantity $\sigma (pp\rightarrow Z') B(Z'\rightarrow
\ell^+\ell^-) \Gamma_{Z'}$ would  yield the information on an overall strength
of
the $Z'$ gauge coupling\cite{CL}\ .
 Here $\sigma (pp\rightarrow Z')$ is the  total cross-section
and $ B(Z'\rightarrow
\ell^+\ell^-)$ the branching ratio for the $\ell^+\ell^-$ final state channel.

 The values of the couplings  (4)  for typical models are given in Table II.
 Note that the couplings  in Eq.(4), probed by  the $LHC$, do not determine the
 couplings in Eq.(2) uniquely. In particular,
determination of $\gamma_L^\ell$, $
\tilde U$ and $\tilde D$  (the three out of four couplings most easily
measurable at the $LHC$) would yield  an eight-fold ambiguity
for the corresponding three
 couplings in  Eq.(2).
Table III  exhibits this two-fold ambiguity for each of the
  $P_V^\ell$ and $P_R^{u,d}$  couplings;   only the first entry is the
actual value of the corresponding coupling in the particular model.

\begin{table}
\caption{%
The  value of the couplings (defined  in Eqs.(2) and (3))
and  statistical error-bars as determined from the probes defined
in Sect. III for  the $NLC$ (c.m. energy $\protect\sqrt s = 500$ GeV and
integrated luminosity ${\cal L}_{int}=20\,\hbox{fb}^{-1}$). The models
are defined  in Sect. II  and  $M_{Z'} = 1$ TeV. 100\%\  heavy flavor
tagging efficiency  and 100\%\ longitudinal polarization of the
electron beam  is assumed for the first set of error-bars, while the
error-bars in   parentheses are for the probes without
polarization.}

\begin{tabular}{c|cccc}
\multicolumn{1}{c}{} & $\chi$ &
$\psi$ & $\eta$ &
{$LR$}\\
\tableline
$P_V^\ell$ & $2.0\pm0.08\,(0.26)$ & $0.0\pm0.04\,(1.5)$ &
$-3.0\pm 0.5\,(1.1)$ & $-0.15\pm 0.018\,(0.072)$  \\
$P_L^q$ & $-0.5\pm 0.04\,(0.10)$ &  $0.5\pm0.10\,
(0.2)$ &  $2.0\pm0.3\,(1.1)$ & $-0.14\pm
 0.037\,(0.07)$
\\
$P_R^u$ & $-1.0\pm0.15\,(0.19)$ & $-1.0\pm0.11\,
(1.2)$ &  $-1.0\pm0.15\,(0.24)$ & $-6.0\pm1.4\,
(3.3)$\\
$P_R^d$ & $3.0\pm0.24\,(0.51)$ & $-1.0\pm0.21\,(2.8)$
& $0.5\pm0.09\,(0.48)$ & $8.0\pm1.9\,(4.1)$\\
$\epsilon_A$ & $0.071\pm0.005\,(0.018)$
& $0.121\pm0.017\,(0.02)$ &
 $0.012\pm0.003\,(0.009)$ & $0.255\pm0.016\,(0.018)$
\\
\end{tabular}
\end{table}

\begin{table}
\caption{
\protect\cite{ACLII}\ Values of the  couplings (4) probed directly
at the $LHC$.  The statistical error-bars indicate how well these couplings
can be measured at the $LHC$ (c.m. energy
$\protect\sqrt s=16$ TeV and integrated
luminosity ${\cal L}_{int}=100\,\hbox{fb}^{-1}$)  for the typical
models  with $M_{Z'}=1$ TeV. }
\begin{tabular}{c|cccc}
\multicolumn{1}{c}{} & $\chi$
& $\psi$ & $\eta$ &
\multicolumn{1}{c}{$LR$} \\
\hline
$\gamma_L^\ell$ & $0.9\pm0.018 $ & $ 0.5\pm0.03 $ &
$0.2\pm0.015 $ & $0.36\pm 0.007 $ \\
$\gamma_L^q$ & $ 0.1 $ & $ 0.5 $ &
$0.8 $ & $ 0.04$ \\
$\tilde U$ & $ 1\pm0.18 $ & $ 1\pm0.27 $ &
$ 1\pm0.14 $ & $ 37\pm8.3$ \\
$\tilde D$ & $9\pm0.61 $ & $ 1\pm0.41 $ &
$ 0.25\pm0.29$ & $65\pm14$ \\
\end{tabular}
\end{table}

\begin{table}
\caption{
Values of three (out of four) couplings (2) which  are
probed (indirectly) at the $LHC$ (see Eq.(5), which relates the
couplings (2) to  those
directly  probed by the $LHC$).  The error-bars indicate how well these
couplings
 can be measured at the $LHC$ (c.m. energy $\protect\sqrt s=16$ TeV and
integrated
luminosity ${\cal L}_{int}=100\,\hbox{fb}^{-1}$)  for the typical
models  with $M_{Z'}=1$ TeV. There is a two-fold ambiguity for
each of the couplings. Only the first number corresponds to the actual value of
the
coupling of the particular model.}
\begin{tabular}{c|cccc}
\multicolumn{1}{c}{} & $\chi$
& $\psi$ & $\eta$ &
\multicolumn{1}{c}{$LR$} \\
\hline
$P_V^\ell$ & $ 2\pm 0.15 $ & $ 0\pm 0.03 $ &
$ -3\pm 0.19 $ & $ -0.148\pm 0.007 $ \\
$\ $ & $ 0.5\pm 0.02 $ & $ \infty\pm \infty $ &
$ -0.333\pm 0.021 $ & $ -7\pm 0.36 $ \\
\hline
$P_R^u$ & $ \mp 1\pm 0.09 $ & $ \mp 1\pm 0.14 $ &
$ \mp 1\pm 0.07 $ & $ \mp 6.04\pm 0.68 $ \\
\hline
$P_R^d$ & $\pm 3\pm 0.10 $ & $ \mp 1\pm 0.21 $ &
$ \pm 0.5\pm 0.29 $ & $ \pm 8.04\pm 0.87 $ \\
\end{tabular}
\end{table}

\section{${\bf {\lowercase{e}}^+{\lowercase{e}}^- \rightarrow
{\lowercase{f}}\bar {\lowercase{f}}}$ Observables}

At the $NLC$  the  cross sections  and corresponding asymmetries  in the
two-fermion  final state channels,  $e^+e^- \rightarrow f\bar f$,
 will be measured.  Due to the
interference of the $Z'$ propagator with the photon and the $Z$ propagators
such probes  are sensitive to the four normalized charges in Eq.(2) as
well as  to  the parameter $\epsilon_A$ (Eq.(3)).
The tree-level expressions for such probes  can be
written explicitly in terms of
  seven generalized charges,  which are given in Ref. \cite{DLRSV}\ .

 The estimates
for statistical and systematic errors suggested \cite{DLRSV}\ an
analysis based on the  following  probes:

\begin{equation}
\sigma ^{\ell},\   \ \ R^{had} = \frac {\sigma ^{had}}
{\sigma ^{\ell}},\ \ \ A_{FB}^{\ell}.
\end{equation}
In the case that  longitudinal polarization of the    electron
beam is available  there are additional  probes:
\begin{equation}
A_{LR}^{\ell, had},\ \ A_{LR,FB}^{\ell}\ .
\end{equation}
 Here $\sigma$, $A_{FB}$, $A_{LR}$ and $A_{LR,FB}$  refer to the
corresponding cross sections, forward-backward asymmetries, left-right
(polarization)
asymmetries and left--right--forward--backward asymmetries, respectively.
 The  superscripts $\ell$ and $had$
refer to all three leptonic channels (considering only
$s$-channel exchange for electrons) and  to all
hadronic final states, respectively.
The above  quantities help to distinguish among different
models\cite{DLRSV}\ ;  however,  they do not yield  information on all the
 $Z'$ couplings. In particular $\sigma^{\ell}$ and $A_{FB}^{\ell}$ probe
$\epsilon_A$  and the magnitude of $P_V^\ell$, but not  its
sign\cite{FOOTA}\ . On the other hand, $R^{had}$
 provides  additional information on one linear
combination of the normalized  quark couplings.
If polarization is available, $A_{FB}^{\ell}$ and
$A_{LR,FB}^{\ell}$  are excellent probes for $P_V^\ell$ (including its sign),
while
$A_{FB}^{had}$ yields  information on another linear combination of the quark
couplings. See  Table IV for the approximate dependence of the above probes on
the couplings.

LEP analyses show that
$e^+e^-$ colliders allow for an efficient tagging of
charm and bottom final states\cite{LEP}\ . Eventually, top events
will also  be easily identifiable  at the future $e^+e^-$ colliders.  We
therefore
assume  that at the $NLC$  an efficient tagging of the heavy flavors ($c,b,t$)
 would be available. This in turn provides an additional set of
observables:
\begin{equation}
R^{f}=\frac{\sigma^f}{\sigma^\ell},\ \ A^{f}_{FB} \ ; \ f=c,b,t\
,\end{equation}
 and with polarization available:
\begin{equation}
A^{f}_{LR}\ \ A^{f}_{LR,FB}\ ;\ f=c,b,t\ ,
\end{equation}
 where the superscript refers to the corresponding
 heavy flavors.
These  additional probes   would in turn allow for a complete  determination of
 the $Z'$ gauge  couplings to ordinary fermions, giving the assumptions of
family universality, $[Q', T_i]=0$, and neglect of $Z-Z'$ mixing (see Sect.
II).

To illustrate quantitatively the sensitivity of the above probes
for the $Z'$ couplings, we display
 the  explicit dependence
 on  the couplings (2) and $\epsilon_A$ (Eq.(3)) in Table IV.
(Table II in Ref. \cite{ACLII}\ provides
analogous expressions for the probes at the $LHC$.) The expressions are at
tree-level,
evaluated to ${\cal {O}} (\epsilon_A)$, only.
In Table IV we neglect fermion masses ($m_f^2 \ll s$),
implying $\sigma ^e = \sigma ^{\mu} =
\sigma ^{\tau} \equiv \sigma ^\ell$
(only $s$-channel exchange is considered for electrons),
$\sigma ^c = \sigma ^t \equiv \sigma ^u,\
\sigma ^b \equiv \sigma ^d$,
and similarly for the  corresponding asymmetries.
(Obviously, neglecting the top mass may be not a good
approximation.)
For $M_{Z'}\sim 1$ TeV, $\epsilon_A$ is sufficiently small, so that
 the use of these expressions  versus the exact Born approximation expressions
changes   the numerical results only  by a few \%. (The numerical  results in
Tables I, V, and VI and the Figures use the exact Born approximations,
including $m_{t}=150$  GeV effects.)

\begin{table}
\caption{
Tree level expressions, correct to ${\cal {O}} (\epsilon_A)$, for
 the total   cross sections, $\sigma ^{f}$,
the forward-backward asymmetries, $A^{f}_{FB}$,
and the corresponding polarized asymmetries,
$A^f_{LR}$ and $A^f_{LR,FB}$  Here,  $ f=\ell,u,d$  are the  final state
flavors.
We neglect fermion masses, $m_f^2 \ll s$, and take the weak
mixing angle $\sin^2\theta _W = 0.23$.
The couplings are defined in Eqs.(2) and (3).
$\sigma _0 = \frac {4\pi \alpha ^2}{3s}$ is the point-like
QED cross section for muon pair production, $\alpha=\frac{1}{128} $ the
electromagnetic coupling constant, and $N_c=3$ is the number of colors.
}
\begin{tabular}{rc}
{$\sigma ^\ell=$}&{ $\sigma _0
[1.140 - (0.500 P^\ell_V\ ^2
+ 0.029 P^\ell_V + 0.183) \epsilon_A ]
$ } \\
{$\sigma ^u =$}& {$\sigma _0 N_c
[0.614 + (0.253 P^q_L
+ 0.354 P^\ell_V P^q_L - 0.112 P^q_L P^u_R
+ 0.324 P^\ell_V P^q_L P^u_R )\epsilon_A] $} \\
{$\sigma ^d = $}&{$\sigma _0 N_c [0.323 - (0.309 P^q_L
+ 0.191 P^\ell_V P^q_L - 0.056 P^q_L P^d_R
+ 0.162 P^\ell_V P^q_L P^d_R )\epsilon_A] $ }\\
\hline
{$A^\ell_{FB} =$} &{$0.483 + (0.091 P^\ell_V\ ^2
- 0.007 P^\ell_V - 0.251 )\epsilon_A$}\\
{$A^u_{FB} =$} &{$0.614 + (0.179 P^q_L
- 0.045 P^\ell_V P^q_L - 0.284 P^q_L P^u_R
- 0.187 P^\ell_V P^q_L P^u_R )\epsilon_A$ }\\
{$A^d_{FB} =$}&{$0.634 +(0.163 P^q_L
-0.343 P^\ell_V P^q_L +0.266 P^q_L P^d_R
+0.188 P^\ell_V P^q_L P^d_R )\epsilon_A$ }\\
\hline
{$A^\ell_{LR} =$} &{$0.070 + (0.018 P^\ell_V\ ^2
- 0.598P^\ell_V - 0.002 )\epsilon_A$}\\
{$A^u_{LR} =$}&{$ 0.348 + (0.433 P^q_L
+0.211P^\ell_V P^q_L +0.591 P^q_L P^u_R
-0.366 P^\ell_V P^q_L P^u_R )\epsilon_A $ }\\
{$A^d_{LR} =$}&{$0.619 + (0.001 P^q_L
-0.591 P^\ell_V P^q_L-0.609 P^q_L P^d_R
+0.484P^\ell_V P^q_L P^d_R )\epsilon_A $} \\
\hline
{$A^\ell_{LR,FB} =$}& {$0.053 + (0.013P^\ell_V\ ^2
- 0.449P^\ell_V - 0.001 )\epsilon_A$}\\
{$A^u_{LR,FB} = $}&{$0.175 + (0.237P^q_L
+ 0.332 P^\ell_V P^q_L + 0.169 P^q_L P^u_R
- 0.488 P^\ell_V P^q_L P^u_R )\epsilon_A $} \\
{$A^d_{LR,FB} =$}&{$ 0.476- (0.262 P^q_L
+ 0.162 P^\ell_V P^q_L +0.213P^q_L P^d_R
- 0.615 P^\ell_V P^q_L P^d_R )\epsilon_A $} \\
\end{tabular}
\end{table}

\section{Determination of ${\bf Z'}$ Couplings at the ${\bf NLC}$}
We now study  how well one can determine the couplings defined in Sect. III
at  the $NLC$.
The effects of a heavy $Z'$   far off-shell are
expected to be small and comparable to the electro-weak
radiative corrections \cite{DLRSV}\ . The latter ones are dominated by initial
state radiation, which can be greatly reduced by applying
a cut on the maximum photon energy to exclude $Z$
production. With such a  cut the tree-level expressions are
a reasonably good approximation to the different observables.
Since our present goal is to explore the sensitivity of the $Z'$ couplings, it
is sufficient to neglect the remaining radiative corrections. Of course, if a
new $Z'$ is actually discovered a realistic
fit should include full radiative corrections as well as
 experimental cuts and detector acceptances.

Throughout  the paper we take  the c.m. energy $\sqrt s = 500$ GeV,
and the integrated luminosity ${\cal L}_{int}=
 20\,\hbox{fb}^{-1}$.
For the analysis we use the probes defined in Eqs.(6-9).
 We assume  100\%  efficiency for heavy
flavor tagging   (probes (8-9)) and  100\%
 longitudinal polarization  of the
initial electron beam for probes (7) and (9). We will,  however,
 also address the case in which the polarization
 and the heavy flavor tagging  efficiency are smaller.
 We include only  statistical errors
for the observables and neglect error correlations for the input
parameters. For this reason, and because we do not include experimental cuts
and detector acceptances our results should be interpreted as a limit on how
precisely  the couplings can be determined for each model for the given
c.m. energy and the integrated luminosity of the $NLC$.
 Realistic fits are expected to give larger uncertainties for the couplings.

In Table V we give the  values of  the probes (6-9)  and their statistical
uncertainties  at the $NLC$ for the typical models.  For comparison,  the
values in the last column correspond to  those of the standard model.
The first row is $\sigma^\ell{\cal L}_{int}$, the number of events
in one $\ell=(e,\mu , \tau)$ channel.

\begin{table}
\caption{
The  values and statistical error-bars for the observables (6-9)
(defined in Sect. III) at  the $NLC$ (c.m. energy $\protect\sqrt s = 500$ GeV
and
integrated luminosity ${\cal L}_{int}=20$
fb$^{-1}$). The models are defined  in Sect. II,  and
$M_{Z'} = 1$ TeV. The first row is the number of events in the
$\ell$ channel. The last column is the standard model ($SM$) prediction. }
\begin{tabular}{c|ccccc}
\multicolumn{1}{c}{} & $\chi$
& $\psi$ & $\eta$ &
{$LR$}&$SM  $ \\
\hline
$\sigma^{\ell}{\cal L}_{int}$ & $ 7850\pm 90$ & $ 8910\pm 90 $ &
$ 8640\pm 90$ & $ 8730\pm90$&$9080\pm 100$\\
$R^{had}$ & $8.73\pm.10 $ & $7.28\pm0.08$ &
$7.69\pm0.09$ & $ 7.18\pm0.08$ &$7.16\pm0.08$\\
$A^\ell_{FB}$ & $0.491\pm0.010$ & $0.451\pm 0.009 $ &
$0.490\pm 0.009 $ & $0.415\pm 0.010 $&$0.483\pm0.009$ \\
\hline
$A^\ell_{LR}$ & $ -0.018\pm 0.008 $ & $ 0.070\pm 0.007 $ &
$0.094\pm 0.008$ & $0.092\pm 0.008 $ &$0.070\pm0.007$\\
$A^{had}_{LR}$ & $0.445\pm0.002 $
& $0.449\pm 0.002 $ &
$0.428\pm 0.002$ & $0.591\pm0.002 $ &$0.448\pm0.002$\\
$A^\ell_{LR,FB}$ & $ -0.013\pm0.008 $ & $0.052\pm0.007 $ &
$0.071\pm 0.008 $ & $ 0.069\pm0.008 $&$0.053\pm0.007$ \\
\hline
$R^c$ & $1.83\pm0.03 $ & $1.71\pm0.02$ &
$1.72\pm0.02 $ & $1.58\pm0.02 $&$1.62\pm0.02 $ \\
$R^b$ & $1.17\pm0.02$ & $0.81\pm0.01 $ &
$0.93\pm0.01 $ & $0.88\pm0.01 $&$0.85\pm 0.01$ \\
$R^t$ & $1.57\pm0.02 $ & $1.43\pm0.02 $ &
$1.45\pm0.02 $ & $1.38\pm 0.02 $ &$1.37\pm0.02$\\
$A^c_{FB}$ & $0.586\pm0.007$ & $0.641\pm0.006 $ &
$0.614\pm 0.006  $ & $0.544\pm0.007 $&$0.615\pm0.007 $ \\
$A^b_{FB}$ & $0.567\pm0.009 $ & $0.625\pm0.009$ &
$0.655\pm0.008  $ & $0.542\pm0.010 $&$0.633\pm0.009$ \\
$A^t_{FB}$ & $0.437\pm0.008 $ & $0.490\pm0.008 $ &
$0.465\pm0.008  $ & $0.398\pm0.008  $ &$0.463\pm0.008$\\
\hline
$A^c_{LR}$ & $0.311\pm0.006  $ & $0.337\pm0.005  $ &
$0.300\pm0.006 $ & $0.474\pm0.005 $&$0.347\pm0.005$ \\
$A^b_{LR}$ & $0.627\pm0.006 $
& $0.657\pm0.006 $ &
$0.633\pm0.006 $ & $0.785\pm0.005 $&$0.620\pm0.006$ \\
$A^t_{LR}$ & $0.347\pm0.006 $ & $0.361\pm0.006$ &
$0.339\pm0.006 $ & $0.491\pm0.006  $&$0.368\pm0.006$ \\
$A^c_{LR,FB}$ & $0.111\pm0.006  $ & $0.178\pm0.006  $ &
$0.117\pm0.006 $ & $0.213\pm0.006 $&$0.174\pm0.006$ \\
$A^b_{LR,FB}$ & $0.379\pm0.007  $ & $ 0.472\pm0.007 $ &
$0.458\pm0.007$ & $0.533\pm0.007  $ &$0.477\pm0.007$\\
$A^t_{LR,FB}$ & $0.083\pm0.006  $ & $0.136\pm0.006 $ &
$0.089\pm0.006 $ & $0.156\pm0.006  $&$0.131\pm0.006$ \\
\end{tabular}
\end{table}
\begin{figure}[p]
\iffiginc
\else
Figures 1a-e go here
\fi
\caption{90\%
confidence level   ($\Delta \chi^2=4.6$) contours  for various pairs of
couplings defined  in
Eqs.(2-3)
for the  $\chi$, $\psi$ and  $\eta$  models (the $LR$ model is in a
different region of parameter space)  at the $NLC$ (c.m. energy
$\protect\sqrt{s} = 500 \hbox{GeV}$
and integrated luminosity  ${\cal L}_{int}=20\,\hbox{fb}^{-1}$)
 and  $M_{Z'}=1$ TeV. Lines correspond to
100\%\ heavy flavor tagging efficiency, and 100\%\ longitudinal  polarization
of the electron beam.
Only statistical error-bars for the probes are  used.}
\end{figure}
\begin{figure}[p]
\iffiginc
\else
Figure 2a goes here
\fi
\end{figure}

\begin{figure}[p]
\iffiginc
\else
Figure 2b goes here
\fi
\caption{
90\%\  confidence level ($\Delta \chi ^2 = 6.3)$ regions for
the $\chi ,\psi $ and $\eta $ models  with $M_{Z'}=1$ TeV are plotted  on
Fig. 2a and Fig. 2b for
$P_R^u$ versus $ P_R^d$ versus $P_V^\ell$ at the $NLC$ (c.m. energy
$\protect\sqrt s = 500$ GeV,
integrated luminosity ${\cal L}_{int}$= 20 fb$^{-1}$) and the  $LHC$
($\protect\sqrt s =16$ TeV,
${\cal L}_{int}$= 100 fb$^{-1}$),
respectively. Only statistical error-bars for the probes are included.
 Fig. 2b reflects a few-fold  ambiguity in the
determination of these couplings at the $LHC$.}
\end{figure}
We perform a simulated $\chi^2$  analysis for the couplings of the
 typical models given  in Table I for  $M_{Z'}=1$ TeV.
The resulting
$1 \sigma $ uncertainties    are also given in Table I.
 The first set of error-bars is with polarization (using probes (6-9))
 while the error-bars in parentheses are without polarization (using
probes (6) and (8)).   The $Z'$  charges
 can typically
 be determined to around $10-20\%$. Without polarization
the  error-bars  increase by a factor $2-10$, and thus yield
only marginal information about the quark couplings. The poor determination of
the
couplings for the $\eta$ model is related to the small value of $\epsilon_A$
in this case.
The $\psi$ model has
 particularly  poorly determined couplings without polarization.

In Figs. (1a-1e) the
90\%\ confidence level   ($\Delta \chi^2=4.6$) contours
are plotted for the various pairs of parameters in the
 $\chi$, $\psi$ and  $\eta$  models (the $LR$ model is in a
different region of parameter space) for  $M_{Z'}=1$ TeV. (They should be
compared to
analogous contours  for the couplings (4) at the $LHC$ in Figs. 1 of Ref.
\cite{ACLII}\ .)
The contours correspond to 100\%\ heavy flavor tagging
efficiency  as well as 100\% electron beam
polarization. For the $\eta$  model  deviation from
Gaussian contours is especially noticeable.  Contours in the case without
polarization turn out to be  unstable, thus indicating  marginal diagnostic
power of the $NLC$ without polarization.

In Fig. 2a  90\%\ confidence level ($\Delta \chi^2=6.3$) regions  are given
in a three-dimensional plot of $P_R^u$ versus $P_R^d$ versus $P_V^\ell$  for
the $\chi$, $\psi$ and $\eta$ models (the $LR$ model is in a  different region
of  parameter space). The error-bars are  again statistical and assume 100\%\
 efficiency for the  heavy flavor tagging and 100\%\
polarization for the  electron beam.

\begin{figure}[p]
\iffiginc
\else
Figure 3 goes here.
\fi
\caption{
90\%\  confidence level ($\Delta \chi ^2 = 6.3)$ regions for
the $\chi ,\psi $ and $\eta $ models  with $M_{Z'}=1$ TeV are plotted for
$\tilde U$ versus $\tilde D$ versus $\gamma_L^\ell$ at the  $LHC$ (c.m. energy
$ \protect\sqrt s =16$ TeV and
integrated luminosity ${\cal L}_{int}=20\,\hbox{fb}^{-1}$). Only
statistical error-bars are included.
}
\end{figure}

We also checked  how the  uncertainty for the couplings are affected
in the case of  smaller, say 25\%\ ,   heavy flavor tagging
efficiency (the error-bars on
the probes (8-9)  increase by a factor of 2)  as well  as in the  case that
the electron beam polarization is reduced to, {say},
50\% (the error-bars on the probes (7) and (9) increase approximately by a
factor of $\sim2$
for small asymmetries \cite{FOOTE}\ ).
Increased error-bars on the couplings
are  given in Table VI; the first (second) set of the error-bars
corresponds to 25\% (100\% ) heavy flavor tagging efficiency and 100\%  (50\%
) electron beam polarization. In the first case the  uncertainties
increase primarily  on the quark couplings by a factor of $\sim2$.  It is seen
that
even 25\% tagging or 50\% polarization efficiency  is still very useful.

The diagnostic power of the $NLC$ for the $Z'$  couplings decreases
 drastically for
$M_{Z'}{\lower3pt\hbox{$\buildrel>\over\sim$}} 1$ TeV. {\it E.g.}, for
$M_{Z'}=2$  TeV, the uncertainties for the  couplings in the
typical models are 100\%, and thus a model-independent determination
of such couplings is difficult at the $NLC$.

\begin{table}
\caption{
The  value of the couplings (defined  in Eqs.(2) and (3))
and  1$\sigma$  statistical error-bars with decreased heavy flavor
tagging efficiency and smaller longitudinal polarization of the
electron beam, as determined from the probes
 in Eqs.(6-9) at   the $NLC$ (c.m. energy $\protect\sqrt s = 500$ GeV and
integrated luminosity ${\cal L}_{int}=20\,\hbox{fb}^{-1}$). The models
are defined  in Sect. II,  and  $M_{Z'} = 1$ TeV.  The first (second)
set  of error-bars  corresponds to  25\%\
 (100\%)  heavy flavor tagging efficiency and 100\% (50\%) electron beam
polarization.}
\begin{tabular}{c|cccc}
\multicolumn{1}{c}{} & $\chi$
& $\psi$ & $\eta$ &
{$LR$} \\
\hline
$P_V^\ell$ & $ 2\pm 0.10(0.12)$ & $ 0\pm 0.05(0.08) $ &
$ -3\pm 0.55 (0.68)$ & $ -0.15\pm 0.022 (0.031)$ \\
$P_L^q$ & $ -0.5\pm 0.08 (0.06)$ & $ 0.5\pm 0.16(0.11)$ &
$ 2\pm 0.56 (0.53) $ & $ -0.14\pm 0.07(0.04) $ \\
$P_R^u$ & $ -1\pm 0.29(0.17) $ & $ -1\pm 0.19 (0.19)$ &
$ -1\pm 0.25(0.19) $ & $ -6.0\pm 2.7(1.7) $ \\
$P_R^d$ & $ 3\pm 0.45(0.35) $ & $ -1\pm 0.37(0.31) $ &
$ 0.5\pm 0.16 (0.15)$ & $ 8.0\pm 3.8(2.3) $ \\
$\epsilon_A$ & $ 0.071\pm 0.005 (0.008)$
& $ 0.121\pm 0.018 (0.017)$ &
$ 0.012\pm 0.004(0.005) $ & $ 0.255\pm 0.017(0.017) $ \\
\end{tabular}
\end{table}

\section{Comparison with The Large Hadron Collider }

In the previous  Section we have seen that at  the $NLC$ efficient heavy
flavor tagging  and electron beam polarization allow for a model independent
determination of  all of the  four normalized  $Z'$  couplings to quarks and
leptons  for a typical class of models, provided  $M_{Z'}
{\lower3pt\hbox{$\buildrel<\over\sim$}} 1$ TeV.
It also yields information on the  parameter ${\epsilon_A}$, a
ratio of an overall gauge  coupling  strength and $M_{Z'}$,
for fixed c.m. energy $s$.

On the other hand, at the $LHC$  $M_{Z'}$   and the total
width, $\Gamma _{Z'}$, can be measured well. The magnitude of
 three  ($\gamma _L^\ell ,\
\tilde U,\
\tilde D $) out of four $Z'$  couplings  to fermions can be
well determined\cite{ACLII}\
 at the $LHC$  for a typical class of models  and $M_{Z'}
{\lower3pt\hbox{$\buildrel<\over\sim$}} 2$ TeV.
The fourth,
$\gamma_L^q$, requires a measurement of the branching
ratio $ B(Z'\rightarrow q\bar q)$, which may be possible
 with  appropriate kinematic cuts, excellent dijet mass resolution and
detailed knowledge of the QCD background in the $Z' \rightarrow jet\ \ jet$
channel\cite{HP,QQ}\ .

The analysis for the determination of ($\gamma_L^\ell ,\ \tilde U,\
\tilde D$) has been done in
 Ref. \cite{ACLII}\ . In the main production channel ($pp\rightarrow
Z'\rightarrow \ell^+\ell^-$, $\ell=e,\mu$)
 the  forward-backward asymmetry  and the ratio of
cross sections in different rapidity bins were used. In the four-fermion final
state channels the
 rare decays $Z'\rightarrow W\ell\nu_\ell$ (with the imposed
$m_{T\ell\nu_\ell}>90 $ GeV cut on the transverse mass of the $\ell \nu_\ell$
system) and associated  productions $pp\rightarrow Z'V$ ($V=Z,W$ and
$V=\gamma$,
  with
$p_{T\gamma}\ge 50$ GeV imposed on the photon transverse momentum) were
used. Only   statistical error-bars  for the probes were incorporated.

The couplings were determined for the  CERN $LHC$ (c.m. energy $\sqrt s=16$
TeV,
integrated luminosity ${\cal L}_{int}=100\,
\hbox{fb}^{-1}$)\cite{FOOTD}\
 for a class of typical models
 and $M_{Z'}=1$ TeV.   The results are summarized  in Table II \cite{ACLII}\
. In  Fig. 3 we  also present a three-dimensional plot,  where
90\%\ confidence level $(\Delta \chi ^2 = 6.3)$  regions for
 $ \tilde U$  versus $\tilde D$ versus $\gamma_L^\ell$ are plotted for
$\chi ,\  \psi $ and $\eta $. Note the
clear separation between the models.

The  couplings  in Eq.(4)  that are  probed directly at the $LHC$
 are  not sensitive to the relative
signs of the $Z'$ charges. This in turn implies that
couplings (2), which are observed directly at the $NLC$,
 are probed with a few-fold ambiguity at the $LHC$.
In Table III we collect the errors
expected at the   $LHC$ for the three couplings  $P_V^{\ell}$, $P_R^u$ and
$P_R^d$.
 We again choose the typical models and $M_{Z'}=1$ TeV.
There is   an eight--fold ambiguity in determination of these couplings;
only the first value of $P_V^{\ell}$, $P_R^u$ and $P_R^d$
corresponds to the actual values of the typical models.
Note, however, that the  error-bars are
typically by a factor of $\sim 2$,   smaller than those at  the $NLC$ (compare
Tables  I  and
III).

In Fig. 2b we plot   90\%\ confidence level $(\Delta \chi ^2 = 6.3)$ regions
for
the $\chi ,\psi $ and $\eta $ models  as
$P_R^u$ versus  $ P_R^d$  versus $P_V^\ell$ at the $LHC$.
While  the error-bars are small,  the figure displays a few-fold ambiguity for
 the value
of the couplings (2) (additional ambiguities are off the scale of the plot).
 At the $NLC$ the
error-bars are on the average
larger, but the   ambiguity in the value of the couplings is now removed.
Thus,  the $LHC$ and the $NLC$ are complementary and together
have a potential to uniquely determine
the couplings with small error-bars .

\section{Conclusions }

We have explored  the diagnostic power of the
$NLC$  (c.m. energy $\sqrt s=500$ GeV, integrated luminosity
${\cal L}_{int}=20\,\hbox{fb}^{-1}$) for a model independent determination of
$Z'$ couplings.
The analysis showed that  efficient heavy flavor tagging
 and longitudinal polarization of the
electron beam provide probes in the two-fermion final state channels, which are
sensitive
 to  the  magnitude as well as the relative signs of  {\it all} the $Z'$
charges
to quarks  and leptons.  For $M_{Z'}
{\lower3pt\hbox{$\buildrel<\over\sim$}} $ 1 TeV, such
couplings would  be determined
to about $10-20$ \%  for a class of typical models.
If the  polarization were not    available,
the determination of the  $Z'$  couplings  would  be  marginal, since the
error-bars
increase by a factor of $2-10$.  Without heavy flavor tagging very
little  can be learned about the quark couplings.

 We  took
into account   the tree-level  expressions for the probes
with  their statistical errors, only.
In addition, we used optimistic, though not unreasonable, assumptions for
the  heavy flavor tagging  efficiency and
the  electron beam polarization. The analysis is thus useful
 for gaining  qualitative information
 on the diagnostic power  of the $NLC$ for $Z'$ couplings.
If a new $Z'$ were known to exist, a realistic
fit should include full radiative corrections,
experimental cuts and detector acceptances, systematic errors and error
correlations. It is expected that
in this case the  error-bars for the couplings would increase.

In the second part of the paper we compared the diagnostic power of
 the $NLC$  with the  $LHC$.
The  $LHC$ is complementary in nature; while
it primarily allows for the  determination  of  the magnitude
of three out of four normalized couplings only,
the corresponding errors  are  typically by a factor of $\sim 2$
  smaller than  those for the
$NLC$  for typical models with $M_{Z'}=1$ TeV.  In addition,
the $LHC$ would measure $M_{Z'}$  directly and would
allow for a determination of an overall strength of the $Z'$ gauge coupling to
fermions. This is  in contrast to  the $NLC$ which, for the fixed c.m.
energy, primarily determines only the ratio of an overall $Z'$ gauge coupling
strength and $M_{Z'}$.

In conclusion,  the analysis  demonstrates the complementarity
of the $NLC$ and $LHC$  colliders, which in conjunction allow for
determination of  $M_{Z'}$, an overall $Z'$ gauge coupling strength as well
as a  unique determination of   {\it all} the quark and lepton charges with
sufficiently small error-bars, provided $M_{Z'}
{\lower3pt\hbox{$\buildrel<\over\sim$}} 1$ TeV.

\acknowledgments

We thank F. Cornet, M. Delfino, J. Liu,
M. Mart\'\i nez and T. Riemann for discussions and D. Benton for  help with
the figures. We are especially grateful to P. Langacker for numerous
discussions and suggestions as well as for  careful reading of the manuscript.
The work was  supported in part
by CICYT under contract AEN93-0615  (F.del A.), the European Union under
contract CHRX-CT92-0004 (F. del A.),
the  U.S. DOE  Grant No. DOE-EY-76-02-3071 (M. C.) and the Texas
National Research Commission Laboratory (M. C.).


\begin{references}
\bibitem{FOOTA} This is due to the property that the photon couplings are
vector-like  and that
 $\hat g_{L1}^\ell\simeq -\hat g_{R1}^\ell$ for  the $\ell$ couplings to
$Z$, which  in turn ensures that
the $\sigma^{\ell}$ and
$A_{FB}^{\ell}$  probe    primarily only  the square of  the $P_V^\ell$
coupling.
\bibitem{FOOTE} The error-bars  $\Delta A_{LR}$ for the polarization
asymmetries
$A_{LR}$ are related to the amount of the beam polarization  $ p $ in the
following way: $\Delta A_{LR}=\sqrt{1-(pA_{LR})^2}/(p\sqrt{N})$. Here $N$ is
the number of the events in the particular channel. Since in the standard
model most of  the asymmetries are small  (see Table V), the corresponding
error-bars scale approximately as $1/p$.
\bibitem{FOOTD} For the  changed projected  c.m. energy $\sqrt s\sim 14$ TeV
 at the $LHC$ and  $M_{Z'}=1$ TeV the cross section in the main
production channel is $\sim$ 30\% smaller  than  for $\sqrt s=16$
TeV.  All the statistical error-bars on the probes then increase by 14\% .
\end{references}
\end{document}